# Electrically reconfigurable nonvolatile flatband absorbers in the mid-infrared with wide spectral tuning range

Romil Audhkhasi[1,†], Virat Tara[1,†], Matthew Klein[2,3], Andrew Tang[1], Rui Chen[4], Shivashankar Vangala[2], Joshua R. Hendrickson[2] and Arka Majumdar[1,3,*]

[1]Department of Electrical and Computer Engineering, University of Washington, Seattle, USA
[2]Sensors Directorate, Air Force Research Laboratory, Wright-Patterson AFB, OH, USA
[3]KBR, Beavercreek, OH, USA
[3]Department of Physics, University of Washington, Seattle, USA
[4]Department of Materials Science and Engineering, Massachusetts Institute of Technology, MA, USA

[*]Corresponding author: arka@uw.edu
[†]These authors contributed equally

**Abstract**
While recent advances in reconfigurable photonics have provided new avenues for manipulating light on the subwavelength scale, on-demand control of infrared absorption remains elusive. Here, we experimentally demonstrate a plasmonic metasurface based on the phase change material $Ge_2Sb_2Te_5$ with in-situ electrically-switchable infrared absorption in the 3 – 5 µm wavelength range. Unlike traditional infrared microstructures based on volatile phase change materials, our device does not require the external stimuli to be continuously applied in order to maintain a given optical state, thus enabling zero static power operation. Furthermore, the 400x deep-subwavelength field localization supported by our device not only allows robust tuning of its spectral response but also makes its absorptivity independent of the angle of incidence, thus enabling a flatband behavior. We conduct switching of our device using rapid thermal annealing and reversible switching using electrical pulses over 26 cycles. Our device provides new avenues for infrared absorption control and serves as a steppingstone for the next generation of mid-wave infrared photonics.

**Introduction**
Recent years have witnessed rapid advances in the development of compact, multi-functional metasurfaces as alternatives to traditional bulky free space optics. Additionally, several approaches to achieve post-fabrication reconfigurability have been leveraged for the realization of tunable spectral filters,[1-3] reconfigurable imagers[4] and novel display technologies.[5] While much of the research effort has been geared towards achieving reconfigurable complex reflection and transmission properties in the visible and near-infrared (NIR) wavelength ranges, on-demand control of absorption at longer wavelengths in the infrared has remained elusive.

Historically, the mid- and longwave infrared (MWIR and LWIR) wavelength range have been of great technological significance for chemical and biological sensing, surveillance, and defense. Reconfigurable photonics has the potential to revolutionize several of the existing application areas in the infrared while enabling the development of new ones. Traditional approaches to achieving reconfigurability such as those relying on the electro-optic[6] and thermo-optic effects[7] or based on incorporating liquid crystals[8] face severe limitations in the MWIR and LWIR. While liquid crystals possess high material loss in these wavelength ranges, thermo-optic effect require constant application of heat resulting in high power consumption and increased thermal noise interfering with the device's optical response.

In recent years, phase change materials (PCMs) have attracted considerable attention from the reconfigurable photonics community due to their ability to exhibit a large refractive index change upon phase transition.[9] Research on tunable photonics in the MWIR and LWIR has largely focused on passive emission control[10-18] by utilizing volatile PCMs such as vanadium dioxide.[19] Volatile PCMs require constant application of heat in order to maintain a particular optical state, and hence suffer from the same limitations as devices based on the thermo-optic effect. On-demand and low-power control of infrared absorption necessitates PCMs with a non-volatile optical response. While several works have investigated reversible optical switching of non-volatile photonic devices using lasers,[20, 21] their dependence on bulky setups and complicated alignment severely limits their practical utility.

Here we propose a plasmonic metasurface based on the non-volatile PCM $Ge_2Sb_2Te_5$ (GST) for in-situ electrical control of infrared absorption in the 3 – 5 μm wavelength range. GST is capable of existing in two stable structural phases at room temperature: crystalline and amorphous, each of which is associated with dramatically different optical properties.[22] Our device leverages deep subwavelength field confinement of 400x within a 10 nm thick GST layer to produce a switchable absorption resonance at a wavelength of 4 μm. This in turn allows the device to achieve a rapid quench rate required for amorphization of the material, thus overcoming a major limitation associated with non-resonant PCM-based photonic devices. We first demonstrate switching of our device with rapid thermal annealing (RTA) followed by reversible tuning over multiple cycles using electrical pulses applied to Si microheaters. We further demonstrate the flatband performance of our device i.e., the robustness of its absorptivity to the angle of incidence, enabled by the strong field confinement associated with the underlying plasmonic structure.

We believe that our non-volatile reconfigurable flatband metasurfaces can have important implications to the fields of infrared sensing and photodetection and can pave the way for several new application areas requiring on-demand, subwavelength control of infrared absorption and emission.

**Results and discussion**

Our proposed device consists of a 10 nm thick layer of GST sandwiched between a 100 nm thick gold stripe grating at the top and a 100 nm thick gold back reflector (Figure 1(a)). The gold stripe grating is periodic in the x direction with a period of 500 nm and the length of the metal stripe in each unit cell is *L*. We calculate the absorption spectrum of the structure using Lumerical FDTD Solutions (see 'Methods' section for details). Figure 1(b) shows the calculated absorption spectrum of a structure with L = 270 nm when GST is in the amorphous and crystalline states. When GST is in the amorphous state, the structure supports a cavity mode at a wavelength of 4 μm which couples to free space resulting in a strong absorption resonance (blue curve). In the crystalline state, the refractive index and loss of GST increases causing a red shift in the resonance wavelength of the cavity mode, a reduction in its peak absorptivity and an increase in its linewidth (red curve). The shift in the absorption peak induced by the phase change of GST causes the absorptivity to reduce from 99.5% in the amorphous state to 1.4% in the crystalline state at a wavelength of 4 µm.

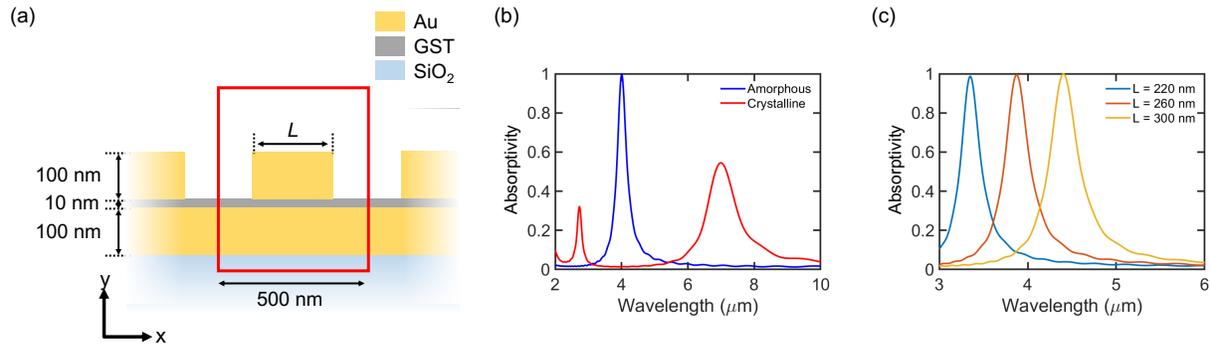

**Figure 1**: (a) Schematic of the proposed device. (b) Absorption spectra of the device when GST is in the amorphous (blue curve) and crystalline (red curve) states. (c) Absorption spectra of three different devices with $L$ = 220, 260 and 300 nm when GST is in the amorphous state.

A salient feature of our structure is the ease of tailoring the location of the absorption peak by changing the length of the grating metal stripe. Figure 1(c) shows the calculated absorption spectra for three different structures with $L$ = 220, 260 and 300 nm when GST is in the amorphous state. We observe that an increase in $L$ causes the absorption resonance to red shift without causing a significant change in the peak absorption and resonance linewidth. Previous studies have shown that the dependence of the peak location on the metal stripe length is linear provided that the length of the metal stripe is not large enough to induce coupling between neighboring unit cells of the device.[23]

To further investigate the nature of the cavity mode supported by our device, we present the on-resonance (wavelength of 4 µm) electric field intensity in a unit cell of the structure with $L$ = 270 nm when GST is in the amorphous state in Fig. 2(a). We observe that the field intensity is strongly localized in the PCM layer under the metal stripe with intensity maxima close to the edges of the stripe. This observation is consistent with previous works that have reported on the existence of localized cavity modes in such structures. For a GST layer thickness of 10 nm and a wavelength of 4 µm, our structure provides 400x field confinement. This not only enables greater flexibility to tailor the spectral response of our absorber as shown in Fig. 1(c) but also makes the spectral response insensitive to the angle of incidence. To illustrate this fact, we present the simulated absorption spectra of the structure in the amorphous state for viewing angles of 0 and 70 degrees in Fig. 2(b). The absorption spectrum at a particular angle of incidence θ is calculated as 1 minus the simulated reflection spectrum of the structure under plane wave illumination at an angle θ. We observe that despite such a large increase in the viewing angle, the absorption peak only exhibits a small blue shift and a slight decrease in the peak absorptivity.

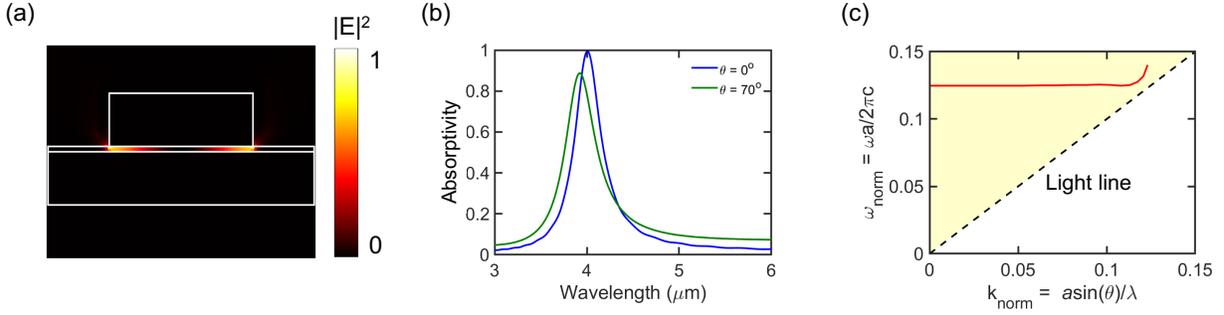

**Figure 2**: (a) Electric field intensity in a single unit cell of a structure with $L$ = 270 nm when GST is in the amorphous state. (b) Simulated absorption spectra of the structure at angles of incidence of 0 and 70 degrees. (c) Optical band diagram of the fundamental cavity mode supported by the structure.

Given the low sensitivity of the cavity mode to the angle of incidence, our proposed device can be regarded as a flatband infrared absorber.[24] To highlight this, Fig. 2(c) presents the optical band diagram along the $x$ direction of the fundamental cavity mode supported by a structure with $L$ = 270 nm in the amorphous state. The band is generated by simulating the structure with the angle of incidence varying from 0 to 80° in increments of 5° (see supporting information for angle-dependent absorption spectra over the full range of angles). The resonance frequencies of the fundamental cavity mode are plotted in normalized units defined in terms of the unnormalized frequencies $\omega$ and unit cell period $a$. Similarly, the component of the wavevector parallel to the direction of the grating ($x$ direction) is expressed in normalized units defined in terms of the viewing angle $\theta$ and the unit cell period $a$. We observe that the cavity mode's optical band is nearly flat upto $\theta$ = 65°, providing further evidence for the insensitivity of the proposed device to a large range of incidence angles.

Next, we fabricate the device using e-beam lithography (see the 'Methods' section for details). Figure 3(a) shows the optical microscope image of the fabricated device connected to metal pads which enable its thermal switching. Figure 3(b) presents a zoomed-in scanning electron microscope image of the device showing the top gold grating. We note that since the gold grating is periodic in only one direction, the spectral response of the device exhibits an absorption peak associated with the fundamental cavity mode only when the illumination is polarized along the direction of the periodicity. Figure 3(c) shows the measured spectral response of three different devices with $L$ = 220, 260 and 300 nm with the period $p$ fixed at 500 nm (see 'Methods' section for details). We observe that an increase in $L$ causes a redshift in the absorption peak consistent with the numerically predicted response presented in Fig. 1(c). We note that the kink in the presented absorption spectra close to a wavelength of 4.2 μm corresponds to the $CO_2$ absorption peak.

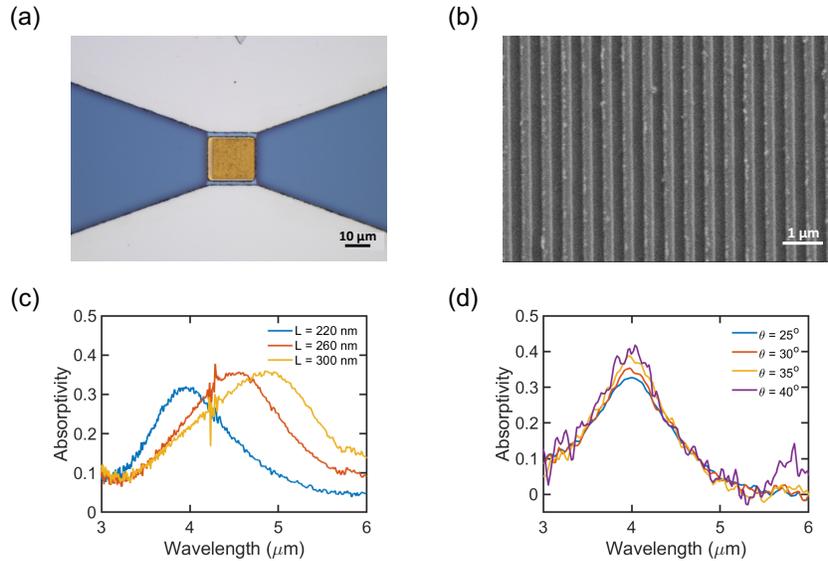

**Figure 3:** (a) Optical microscope and (b) scanning electron microscope image of the fabricated device with a lateral span of 17x17 µm². (c) Measured absorption spectra of devices with varying lengths of the gold grating metal stripes. (d) Measured absorption spectra of the device with $L$ = 270 nm for different angles of incidence.

Figure 3(d) shows the absorption spectra of a device with $L$ = 270 nm for incidence angles varying from 25 to 40°. We observe that an increase in the angle of incidence does not cause any change in the resonance wavelength of the absorption peak, further validating the angle insensitivity of the proposed device. We note that angles of incidence smaller than 25 degrees could not be investigated due to constraints of our experimental setup while those above 40 degrees showed reduced signal-to-noise ratio due to a smaller amount of incident light hitting the detector.

Finally, we experimentally investigate the switching performance of our device. We first perform rapid thermal annealing (RTA) on a 100x100 µm² device) to switch it from its initial amorphous state to the crystalline state. The measured absorption spectra of the device in the two states are presented in Fig. 4(a). As expected from Fig. 1(b), the device exhibits an absorption peak close to a wavelength of 4.5 µm in the amorphous state while having very low absorptivity over the entire 3 – 6 µm wavelength range in the crystalline state. We note that absorptivity values less than 0 in the 3 – 4 µm wavelength range are a result of improper normalization.

Next, we perform multi-cycle electrical switching of a device with an area of 17x17 µm². We note that this device area was chosen to achieve reliable switching of the GST layer over multiple cycles. We note that in this study, we perform switching over only 26 cycles due to the large data acquisition time required for each cycle (see supporting information for details). Each switching cycle comprises of an amorphization and a crystallization pulse. The amorphization pulse is the same for all 26 cycles with a peak voltage of 24V, a rise time of 10ns, width of 300ns and fall time of 10ns. The crystallization pulse for the first 21 cycles has a peak voloage of 5V, a rise time of 1µs, width of 50µs and a fall time of 50µs. We note that a lower amplitude and longer trailing edge pulse can help in crystallizing the GST layer by restraining its critical cooling rate.[25] In the last 5 cycles, we demonstrate the fast switching performance of the device by modifying the crystallization pulse to have a peak voltage of 19V, rise time of 21ns, width of 300ns and fall time of 21ns. We note that for this set of measurements, the SiO$_2$ substrate below our device (see Fig. 1(a)) is replaced with a doped Si-on-insulator substrate. It not only acts as a microheater to which electrical pulses are applied via metal pads (see 'Methods'), but also allows us to achieve uniform heating of the device (see supporting information for details),

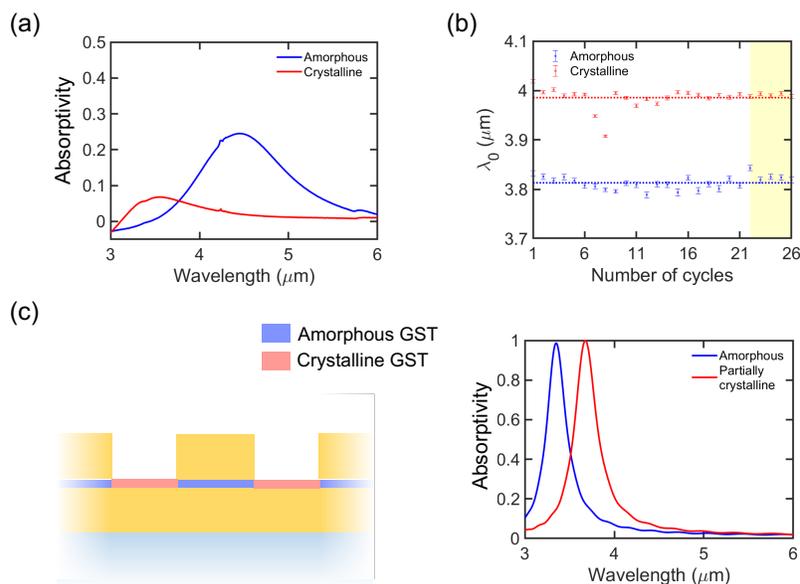

**Figure 4:** (a) Measured spectral response of a 100x100 μm$^2$ device under RTA when GST is in the amorphous (blue curve) and crystalline (red curve) states. (b) Experimentally measured variation of the resonance wavelength of the cavity mode supported by the device in the amorphous and crystalline states over 26 switching cycles. The yellow shaded region represents the fast-switching regime. (c) (left panel) Schematic showing partial switching of the device's GST layer, and (right panel) corresponding spectral response for a device with $L$ = 220 nm.

Figure 4(b) presents the change in the resonance wavelength of the device over 26 cycles as the GST switches between the amorphous (blue dots) and crystalline (red dots) states. The dashed red and blue lines correspond to the mean resonance wavelengths in the two device states. We observe that the resonance wavelengths of the device in the two states stay consistent over all 26 cycles, with a smaller inter-cycle variation for the amorphous state compared to the crystalline state. We note that the observed resonance wavelength change between the two states is about 200 nm which is smaller than what we observe in the case of switching using RTA. Previous studies have shown that the chemistry of the capping layer influences the crystallization kinetics of the PCM layer which in turn causes variations in the crystallization temperature and either promotes or impedes nucleation within the PCM.[26] We hypothesize that the presence of a thick gold grating in the structure introduces non-uniformities in the switching condition of the GST layer. This makes it difficult to switch the regions of the GST layer covered by the gold grating with the peak voltage values used in this study.

To test the above hypothesis, we calculate the absorption spectrum of the structure for the case when the entire GST layer is in the amorphous state and when only the regions of the layer capped by the god grating are in the amorphous state (left panel of Fig. 4(c)). The resulting spectra are displayed on the right panel of Fig. 4(c). We observe that in the case of partial crystallization, the absorption peak is red-shifted by approximately 300 nm with respect to the peak for the amorphous state. This is consistent with our observation in the reversible switching experiments and validates our hypothesis of the non-uniform switching of the GST layer. We note that while using higher peak voltages for reversible switching will allow for full crystallization of the GST layer, it will reduce the device's endurance to multiple switching cycles. A more feasible solution to this issue involves optimizing the gold grating thickness to achieve full crystallization of the GST layer without sacrificing the device's optical functionality.

**Conclusion**

In conclusion, we developed a GST – based plasmonic metasurface with electrically reconfigurable infrared absorption in the 3 – 5 µm wavelength range. Our device exhibits a large change in its absorptivity at a wavelength of 4 µm when GST switches from the amorphous to the crystalline state owing to a red shift in the fundamental cavity resonance. The localized nature of the cavity mode not only allows convenient tailoring of the device's wavelength-dependent absorptivity by appropriately patterning its top metallic layer but also makes it invariant to the angle of incidence. We experimentally validated the flatband behavior of the device and demonstrated the tailoring of its absorption response by modifying the length of the metal stripes in the top grating layer. Furthermore, we demonstrated switching of the device using rapid thermal annealing and reversible, in-situ electrical switching via doped silicon heaters. Under reversible switching, the device exhibited good endurance over 26 switching cycles, with the last 5 cycles corresponding to fast crystallization using a 300ns pulse width.

We note that the current device architecture can be modified to produce a two-dimensional flatband in the *x-z* plane. To achieve this, one may replace the top, one-dimensional metal stripe grating with a two-dimensional lattice of nanodisks to enable both angle and polarization invariance. The absorption change exhibited by the device upon reversible switching can potentially be increased by reducing the thickness of the patterned metallic layer. Additionally, future work may also develop a more accurate modeling of the structural imperfections incurred during device fabrication to optimize the absorptivity contrast at the wavelength of interest.

Despite the above limitations, our device provides new avenues for the multi-attribute control of infrared absorption and emission and serves as a steppingstone for the next generation of photodetectors, hyperspectral imagers, and chemical sensors.

**Methods**

**Simulation approach**

The structure is illuminated with a plane wave normally incident along the *y* axis (see Fig. 1(a)) and reflection is recorded by placing a power flux monitor above it. We use periodic boundary conditions at the edges of the unit cell parallel to the *y* axis and perfectly matched layer boundary conditions along those parallel to the *x* axis. The absorptivity of the structure is calculated as 1 – reflectivity. We neglect transmission through the structure as the thickness of the back reflector is greater than the skin depth of gold in this wavelength range.

**Device fabrication**

The device for RTA switching and variable incident angle measurements was fabricated on a 500µm thick quartz substrate. A blanket film of 100 nm thick gold (Au) was deposited using sputtering (Evatec LLS EVO Sputter System). 10 nm of Cr was used as the adhesion layer which was also deposited using sputtering. Next, 10 nm thick film of GST was DC sputtered (Lesker Lab 18), followed by a lithography step to make the Au gratings. PMMA 495K A6 was spin coated and exposed using 100kV E-beam lithography (JEOL-6300FS) to prepare the sample for Au lift off. Au deposition was carried out using e-beam evaporation (CHA SEC-600) and lift off was performed in Acetone overnight. Device fabrication for reversible switching largely followed the above steps with some modifications. Firstly, instead of quartz, a 220 nm thick silicon on insulator (SOI) substrate was used. The sample was doped using arsenic and later annealed to provide a doping concentration of roughly < 1e20 cm$^{-3}$. We note that the back reflector Au was not used as the heater because of its very low resistance. Ti / Pt metal pads of thickness 20 nm / 150 nm were deposited using sputtering (Evatec LLS EVO Sputter System). To avoid short circuiting between the back reflector Au and doped Si, 200 nm thick Al$_2$O$_3$ was deposited between them using e-beam evaporation (CHA SEC-600). Finally, the device was capped using 20 nm thick Al$_2$O$_3$ deposited using atomic layer deposition at 150C (Picosun R200) to avoid oxidation of the PCM during its switching. The final device resistance is ~151 ohms.

**Equipment used for switching of the device**

Switching using RTA was performed at 275°C using Allwin21 Corp AW 610. Electrical switching was carried out using a pair of probe positioners (Cascade Microtech DPP105-M-AI- S). Pulses for electrical switching were applied using a function generator (Keysight 81160A) attached to a power amplifier (Cleverscope CS1070) with 50 ohm output. All the voltages mentioned in the manuscript are applied voltages.

**Absorption spectrum measurements**

The spectrum of the device was recorded using Bruker HYPERION FT-IR. A linear polarizer was added to polarize the incident light from the source while a 15x objective with NA 0.4 was used to focus it on to the sample. The variable angle measurements were performed with a spectroscopic ellipsometer (IR Vase, J. A. Woolam). The ellipsometer was run in transmission mode which acts as a FTIR spectrometer with a SiC globar light source. The data were taken with a ~1 mm aperture in front of the sample to ensure only light that passes through the device was measured. Angular measurements were taken by rotating the sample about its central axis using an automated rotation stage. The device signal was normalized by the transmission spectra with only the aperture in place. Data was taken with a spectrometer resolution of 8 cm$^{-1}$ and averaged over 500 scans per angle.

<u>**Supporting information**</u>

Simulated angle-dependent absorption spectrum of the device, thermal simulations of the micro-heater, fabrication process flow for reversible switching of the device and the reason for performing device reversible switching over only 26 cycles.

**Acknowledgements**

The work is supported by DARPA-ATOM program. Part of this work was conducted at the Washington Nanofabrication Facility / Molecular Analysis Facility, a National Nanotechnology Coordinated Infrastructure (NNCI) site at the University of Washington with partial support from the National Science Foundation via awards NNCI-1542101 and NNCI-2025489. Research performed by M.K. at the Air Force Research Laboratory was supported by contract award FA807518D0015. J.R.H. acknowledges support from the Air Force Office of Scientific Research under award number FA9550-25RYCOR006.

# Supporting information for "Electrically reconfigurable nonvolatile flatband absorbers in the mid-infrared with wide spectral tuning range"

Romil Audhkhasi[1,†], Virat Tara[1,†], Matthew Klein[2], Andrew Tang[1], Rui Chen[1], Joshua Hendrickson[2] and Arka Majumdar[1,3,*]

[1]Department of Electrical and Computer Engineering, University of Washington, Seattle, USA
[2]Air Force Research Laboratory, Wright-Patterson AFB, OH, USA
[3]Department of Physics, University of Washington, Seattle, USA
[*]Corresponding author: arka@uw.edu

[†]These authors contributed equally

Number of pages: 3
Number of figures: 4

## S1: Simulated angle-dependent absorption spectrum of the device

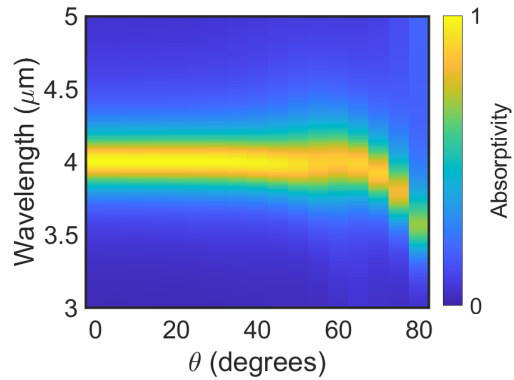

**Figure S1:** Simulated absorption spectrum of a structure with *L* = 270 nm when GST is in the amorphous state.

## S2: Thermal simulations of the micro-heater

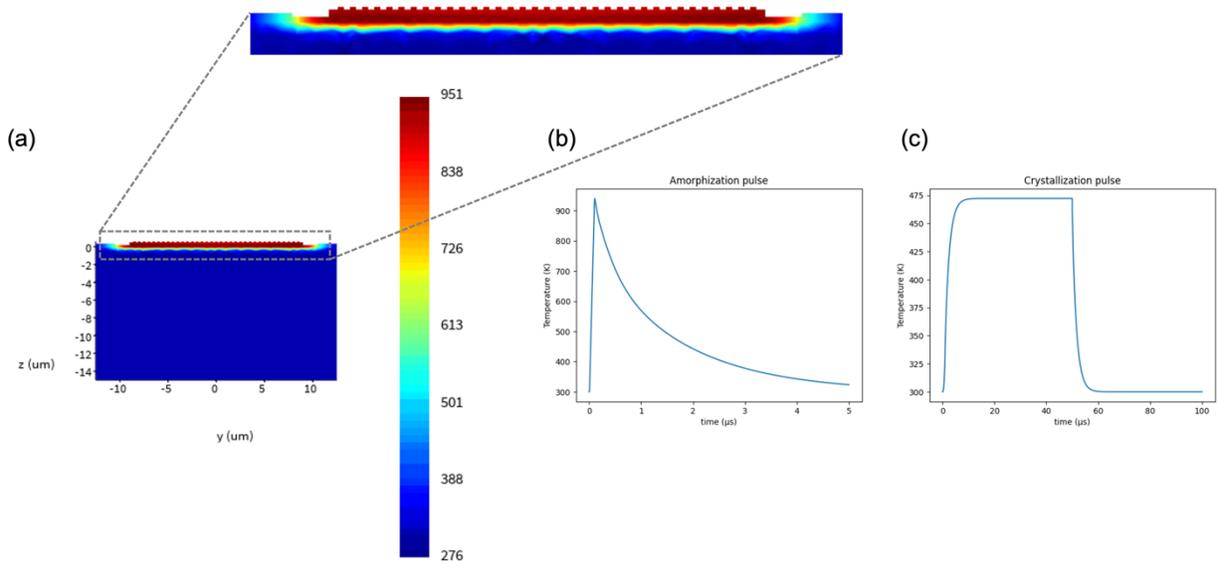

**Figure S2:** 2D thermal simulation performed using the Ansys Lumerical HEAT solver. (a) Heat distribution (side profile) for an 14V 100ns amorphization pulse. (b) Heat at the center of the micro-heater (Temperature v/s time) for amorphization pulse of 14V 100ns pulse width. Using this pulse, the temperature of the micro-heater goes above the melting point of GST (~616°C) and then comes down below the glass transition temperature (~155°C) rapidly. (c) Heat at the center of the heater for crystallization pulse of 2V 50µs pulse width which is enough to generate heat above the glass transition temperature of GST (~155°C).

## S3: Fabrication process flow for reversible switching of the device

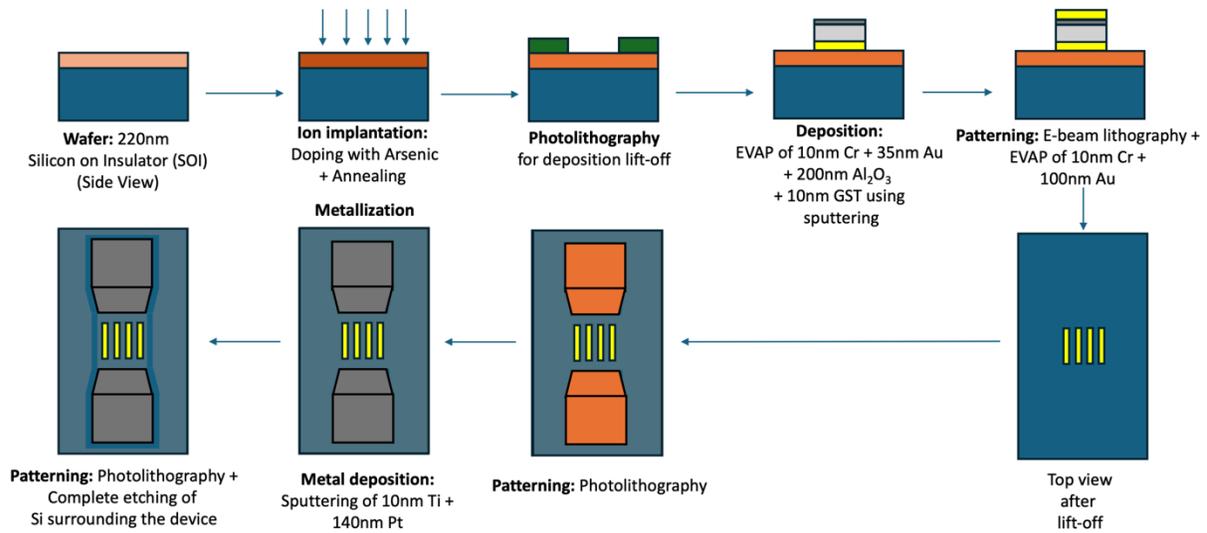

**Figure S3:** Step wise fabrication process flow used to fabricate devices for reversible switching of the infrared absorber.

## S4: Reason for performing device reversible switching over only 26 cycles

We note that in this study, we use FTIR spectroscopy to record the spectral response of the designed absorber, with 256 scans per measurement. This results in a large data acquisition time for each cycle in the reversible switching case, making it challenging to perform switching over a large number of cycles. We note that the device is able to endure the 26 switching cycles, as evident from the similarity of the optical microscope images shown in Fig. S4.

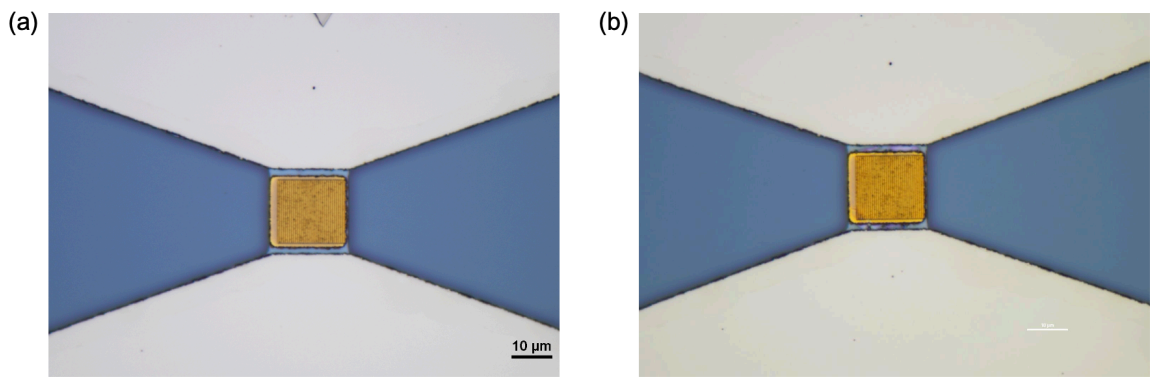

**Figure S4:** Optical microscope images of the device (a) before and (b) after 26 switching cycles. The scale bar is 10 µm for both images.